\documentstyle[preprint,prb,aps,epsf,eqsecnum]{revtex}
\begin{document}
\tightenlines

\title{Bethe's equation is incomplete for the XXZ model at roots of unity  }

\author{Klaus Fabricius
\footnote{e-mail Fabricius@theorie.physik.uni-wuppertal.de}}
\address{ Physics Department, University of Wuppertal, 
42097 Wuppertal, Germany}
\author{Barry~M.~McCoy
\footnote{e-mail mccoy@insti.physics.sunysb.edu}}               
\address{ Institute for Theoretical Physics, State University of New York,
 Stony Brook,  NY 11794-3840}
\date{\today}
\preprint{YITPSB-00-52}

\maketitle

\begin{abstract}

We demonstrate for the six vertex and XXZ model parameterized 
by $\Delta=-(q+q^{-1})/2\neq \pm 1$
that when $q^{2N}=1$ for integer $N\geq 2$ the Bethe's ansatz equations
determine only the eigenvectors which are the highest weights of the
infinite dimensional $sl_2$ loop algebra symmetry 
group of the model. Therefore in this
case the Bethe's ansatz equations are incomplete and further
conditions need to be imposed in order to completely specify the wave
function. We  discuss how the evaluation parameters of the finite
dimensional representations of the $sl_2$ loop algebra can be used
to complete this specification.

\end{abstract}
\section{Introduction}

The study of the eigenvectors and eigenvalues of the Hamiltonian of
 the  XXZ chain with periodic boundary conditions
specified by 
\begin{equation}
H=-{1\over 2}\sum_{j=1}^L (\sigma^x_j\sigma^x_{j+1}+
\sigma^y_j\sigma^y_{j+1}+\Delta \sigma^z_j\sigma^z_{j+1})
\label{ham}
\end{equation}
where $\sigma_j^i$ is the $i$ Pauli spin matrix at site $j$ 
was initiated by Bethe \cite{bethe} for the case $\Delta=\pm 1$ 
in 1931 and has been
studied \cite{orbach}-\cite{yy} for $\Delta \neq \pm 1$ since 1959. 
A major result
of these studies is that the ground state eigenvalue for any value of
$S^z$ is determined by
an appropriate solution of what is called ``Bethe's equation''
\begin{equation}
\left({\sinh {1\over 2}(v_j+i\gamma)\over \sinh{1\over 2}
(v_j-i \gamma)}\right)^L=\prod_{l=1\atop l\neq j}^{{L\over 2}-|S^z|}
{\sinh{1\over 2}(v_j-v_l+2i\gamma)\over\sinh{1\over 2}(v_j-v_l-2i\gamma)}
\label{beq}
\end{equation}
where we use
\begin{equation}
-\Delta=\cos \gamma ={1\over 2}(q+q^{-1}),~~~~~~~~~0\leq \gamma \leq
\pi.
\end{equation}
Here
\begin{equation}
S^z={1\over 2}\sum_{j=1}^L\sigma^z_j
\end{equation}
is a conserved quantum number since the operator in the right hand
side commutes with the Hamiltonian (\ref{ham}) 
The eigenvalues of (\ref{ham}) are
\begin{equation}
E=-{\Delta\over 2}L-2\sum_{j=1}^{{L\over 2}-|S^z|}\left(\cos p_j-\Delta\right)
\label{energy}
\end{equation}
where
\begin{equation}
\cos p=-\cos\gamma+{\sin^2\gamma\over \cosh v -\cos \gamma}.
\label{pdef}
\end{equation}
The corresponding momentum $P$ is obtained from
\begin{equation}
e^{iP}=\prod_{j=1}^{{L\over 2}-|S^z|}{\sinh {1\over2}
(i\gamma+v_j)\over \sinh{1\over 2}(i\gamma-v_j)}.
\label{mom}
\end{equation}

There are many solutions of (\ref{beq}) but it is a simple matter \cite{yy} to
determine the particular solution that leads to the ground state.
However the derivation of the equations is not
restricted to the ground state. Thus there arises the question of
whether or not the totality of solutions of the Bethe's 
equation (\ref{beq}) will
give all eigenvalues of (\ref{ham}). This is referred to as the
completeness problem for the Bethe ansatz equation.

There is a large literature concerning this completeness problem
 \cite{tak2}-\cite{nlk} and
the closely related problems of classifying solutions of Bethe's
 equation in terms of the string \mbox{hypothesis \cite{ts}-\cite{tak}}
and quartets, and wide and narrow pairs \cite{woy}-\cite{bdvv,kz1,kz2}.
 As recently as ref. \cite{nlk} it was stated that ``it still remains
 to be settled whether the Bethe ansatz produces the complete set of
 eigenstates''.

However it is not always straightforward to   interpret precisely what has
 been done. For example in the special case of the XXX  model
 ($\Delta=- 1$) it is shown in ref.\cite{bethe,fadtak2}-\cite{kir1} 
 that
 the $sl(2)$ symmetry of the Hamiltonian groups states in degenerate
 multiplets where,
denoting by $S^z_{\rm
max}$ the maximum value of $S^z$ in the multiplet, there are
$2S^z_{\rm max}+1$ states in the multiplet with the same energy and momentum
and with the values $S^z= S^z_{\rm max},~S^z_{\rm
max}-1,~\cdots,-S^z_{\rm max}.$ These states are all given by Bethe's
equation as long as multiple occupancy of the states with $v_j=\pm
\infty$ is allowed.   
On the other hand as is pointed out in
 ref. \cite{tarvar} these computations are based on the string
 hypothesis of ref. \cite{ts} and in
 ref. \cite{dl,eks,vlad} it is demonstrated 
that there are states 
of the XXX model where this hypothesis fails.

Another example is the study of completeness made in  ref. \cite{lanst1}
where it is stated that when
$|\Delta|<1$ the root of unity case
\begin{equation}
q^{2N}=1
\label{root}
\end{equation}
must be explicitly excluded  for the completeness proof to hold 
because of the occurrence of 
additional degeneracies.
However, in ref. \cite{kirlis} this root of unity case is
precisely the case for which it
is claimed that a combinatorial completeness is proven 
on the basis of assuming a counting given by the string hypothesis
and in ref.\cite{kirlis} there is no explicit
mention made of degenerate multiplets mentioned in
ref. \cite{lanst1}.

The study of the degeneracies of the XXZ model at roots of unity
originates in the much more general studies of Baxter \cite{baxc}- \cite{baxe}
on the XYZ model and has been considered in the more special case of
XXZ by several authors \cite{korep1}-\cite{korep3}.
Recently \cite{dfm} it was shown in the root of unity case 
(\ref{root}) that the Hamiltonian (\ref{ham}) and
the transfer matrix of the related \mbox{six vertex model
\cite{lieba}-\cite{syy}}  have the infinite
dimensional $sl_2$ loop algebra as a symmetry. 
This infinite dimensional symmetry algebra
groups eigenvalues into degenerate multiplets whose consecutive values of $S^z$
differ by $N$ and as an example when $S^z\equiv 0 ({\rm mod}~N)$ the
state with $S^z=S^z_{\rm max}-lN$ has the binomial  multiplicity
${2S^z_{max}/N\atopwithdelims() l}.$  
This $sl_2$ loop algebra symmetry must also lead to consequences for the
 solutions of the corresponding Bethe's equation (\ref{beq}) but
unlike the XXX model with the finite symmetry
algebra of $sl(2)$
multiple occupancy 
of $v_j=\pm \infty$ does not explain this degeneracy. It is the
 primary purpose of this paper to study the relation of the $sl_2$
loop algebra symmetry discovered in ref. \cite{dfm} to the Bethe's
equation (\ref{beq}). 

In order to efficiently study all the solutions of Bethe's equation we
find it most useful to recall that
(\ref{beq}) arises in Baxter's functional
equation solution \cite{baxa}-\cite{baxb},\cite{baxc}-\cite{baxe} 
to the six vertex model
where the (suitably normalized) transfer matrix $T(v)$ of 
the six vertex model satisfies
\begin{equation}
(-1)^{{L\over 2}-|S^z|}T(v)Q(v)=\sinh^L{1\over 2}(v-i\gamma)Q(v+2i\gamma)+
\sinh^L{1\over 2}(v+i\gamma )Q(v-2i\gamma)
\label{feq}
\end{equation}
where $T(v)$ is defined in
terms of the Boltzmann weights $W(\mu,\nu)|_{\alpha,\beta}$ with
$\mu,\nu,\alpha,\beta=\pm 1$ as
\begin{equation}
T(v)={\rm Tr}W(\mu_1,\nu_1)W(\mu_2,\nu_2)\cdots W(\mu_L,\nu_L)
\end{equation}
where 
\begin{eqnarray}
W(1,1)|_{1,1}&=&W(-1,-1)|_{-1,-1}=\sinh{1\over 2}(v+i\gamma)\nonumber
\\
W(-1,-1)|_{1,1}&=&W(1,1)|_{-1,-1}=\sinh{1\over 2}(v-i\gamma)\nonumber
\\
W(-1,1)|_{1,-1}&=&W(1,-1)|_{-1,1}=\sinh i\gamma
\end{eqnarray}
and $Q(v)$ is the  auxiliary matrix introduced by Baxter which satisfies
$[T(v),Q(v)]=[Q(v),Q(v')]=0.$ From these commutation relations and the
fundamental relation that $[T(v),T(v')]=0$ it follows that the matrix
equation (\ref{feq}) also holds for the eigenvalues of $T(v)$ and $Q(v).$
Therefore if we write the eigenvalues of $Q(v)$ in the product form
\begin{equation}
Q(v)=\prod_{j=1}^{{L\over 2}-|S^z|}\sinh{1\over2}(v-v_j)
\label{qfac}
\end{equation}
we find from (\ref{feq}) that 
the zeros $v_j$ are given by (\ref{beq})
as long as the simultaneous vanishing
\begin{equation}
Q(v_j)=Q(v_j+2i\gamma)=Q(v_j-2i\gamma)=0
\label{vanishing}
\end{equation}
does not occur.

In general this is all which is known. However, in the special case
where $S^z=0$ Baxter in equation 101  of ref.\cite{baxc}
gives the following explicit expression for the matrix $Q(v)$ which is
valid for all $\gamma$ 
\begin{equation}
Q(v)|_{\alpha,\beta}=\rho{\rm exp}\left({1\over 4}i(\pi-\gamma)\sum_{1\leq
J<K\leq L}(\alpha_J \beta_K-\alpha_K\beta_J)+{1\over
4}(v-i\pi)\sum_{J=1}^{L}\alpha_J\beta_J \right)
\label{qform}
\end{equation}
where $\alpha_j,\beta_j=\pm 1$ are the eigenvalues of $\sigma^z_j$
at the site $j,$ $\rho$ is a suitable normalizing 
constant and the restriction $S^z=0$ means that $L$ is even and
that
\begin{equation}
\alpha_1+\cdots+\alpha_L=\beta_1+\cdots +\beta_L=0.
\label{qmat}
\end{equation}

The functional equation (\ref{feq}) and the exact expression
(\ref{qform}) for $Q$ with $S^z=0$  will be the basis for
most of our studies of the solutions $v_j$ of Bethe's equation (\ref{beq}).

In order to make our statements precise we will first summarize in sec. 2
what is meant by the string hypothesis of refs. \cite{ts} and
\cite{tak} and formulate procedures for numerically determining the
solutions $v_j$ of the Bethe's equation (\ref{beq}).
In section 3 we will study
these numerical solutions of (\ref{beq}) for
the two values $\Delta =0, -1/2$ and compare with the picture given by
the string hypothesis \cite{ts,tak}  and the counting of
ref.\cite{kirlis}. 
For the case $\Delta=0$ the complete solution was
given long ago by Lieb, Schultz and Mattis \cite{lsm}.
We will see that the $v_j$ which are obtained from the Bethe's ansatz
equation (\ref{beq}) in the limit $\Delta\rightarrow 0$ and from the
exact expression (\ref{qform}) for $Q$ with $S^z=0$ do not
agree with the corresponding $v_j$ of \cite{lsm} whereas the 
state counting results of ref. \cite{kirlis} do agree with  \cite{lsm}.
Similarly for $\Delta=-1/2$ the roots obtained from continuity and
from the
exact expression (\ref{qform}) for $Q$ with $S^z=0$ are 
also different from the string structure posited by
ref. \cite{kirlis}.

It must be stressed, however, that even though  
the numerical results 
we find for $v_j$ do not agree with the results of ref. \cite{lsm} for $\Delta=0$ 
our results do not contradict this paper
because the
eigenvectors being discussed are linear combinations of the  Bethe states
with degenerate eigenvalues which we obtained by continuity. 
We explain this in detail in sec. 4 and
relate this phenomenon to the picture obtained 
from the evaluation representation
decomposition of the $sl_2$ loop algebra. For  
for the case $\Delta=0$ we explicitly compute these evaluation parameters  
using Jordan Wigner
techniques. For both $\Delta=0,-1/2$ we present  an empirical 
relation between the
evaluation parameters computed as roots of the associated Drinfeld
polynomial and some of the numerical solutions of sec. 3 
 
We conclude by remarking that even though for the explicit roots of
unity with 
$N=2,3$ presented in sec. 3 the phenomenon of 
quartets \cite{woy}-\cite{bdvv,kz1,kz2}
does not happen we have in fact found many examples for $N\geq 4$ where
quartet states are present. The existence of these non-string states
has nothing to do with the degeneracies resulting form the $sl_2$ loop
algebra of ref. \cite{dfm} and therefore we have not extended our
examples to cases with $N\geq 4$. 
However these quartets are most
interesting in their own right and for that reason we will discuss them
separately elsewhere.

\section{Formulation}

In this section we make precise what is meant by the string hypothesis
and we outline the procedure we will use to numerically study the
roots $v_j$ of the Bethe's equation (\ref{beq}).

\subsection{The string hypothesis}

The detailed study of the solutions of the Bethe's equation
(\ref{beq}) was begun in 1972 in ref.\cite{ts} where it is
hypothesized that in the limit $L\rightarrow \infty$
the roots $v_j$ form complexes of $n$ values where the
imaginary parts are either
\begin{equation}
\Im v=\cases{(n+1-2k)\gamma& mod $2\pi$\cr
(n+1-2k)\gamma+\pi&mod $2\pi$\cr}
\end{equation}
with $k=1,2,\cdots ,n.$
In other words as $L\rightarrow \infty$ the roots of (\ref{beq}) are
of the form (eqn. 2.9 of ref.\cite{ts})
\begin{equation}
v_{j,k}=\cases{ v^r_j+(n+1-2k)i\gamma +O({\rm exp}(-\delta L))&mod $2\pi$ \cr
      v^r_j+(n+1-2k)i\gamma+\pi i +O({\rm exp}(-\delta L))& mod $2\pi$\cr}
\label{sform}
\end{equation}
where $\delta>0, v_j^r$ is real and $k=1,2,\cdots,n$
The first type of solution is called an $n$ string of positive parity
(denoted as $(n+)$) and the second type of solution is
called an $n$ string of negative parity (denoted by $(n-)$).

The form of solution (\ref{sform}) is universally referred to as the
string hypothesis although some authors include the term 
$O({\rm exp}(-\delta L))$ only implicitly. It is mandatory that these
terms are included for all $n\geq 2$ because if they are not 
the right hand side of (\ref{beq}) will contain explicit  factors of 
zero or infinity.

The next question which arises is which values of $n$ are allowed for
a given value of $\gamma.$ For an irrational value of $\gamma/\pi$ an
infinite number of values of $n$ are allowed when $L\rightarrow
\infty$. However for the values
$\gamma=m \pi/N$ (when the root of unity condition $q^{2N}=1$ holds)
the maximum value of $n$ determined by the formalism of
eqns. (2.12)-(2.14) of ref. \cite{ts} is $n_{max}=N-1.$  

For the purposes of this paper we will  restrict our
attention to the special case $\gamma=\pi/N.$ Then the formalism of ref. 
\cite{ts} gives the following allowed states of strings:
$(1,+),~(2,+),\cdots,(N-1,+)$ and $(1,-).$

There are several assumptions which have been tacitly made in these
 derivations which should be made explicit.
 
First of all there is the assumption which is sometimes made that all the real
parts $v_j^r$ in (\ref{sform}) should be finite. However, 
examples of infinite roots are known for both the XXX and XXZ models
(see for example ref. \cite{sid}) and
 for the XXX model these infinite roots are connected with 
the $sl(2)$ multiplet structure of
the degeneracies.  These infinite roots mean only that the number of finite
roots is less than the assumed value of ${L\over 2}-|S^z|$ which
appears in (\ref{beq}). We do not make this assumption of finiteness
of $v_j^r$ in our definition of string state and thus these
``singular solutions'' do not violate our form of the hypothesis.

There is a more serious problem, however, which first seems to have
 been  made explicit
in 1997 by Takahashi \cite{tak}  who realized that 
the picture given above for $\gamma=m \pi/N$ is not complete
and that strings of length $N$ must also be allowed. We call these strings
complete $N$ strings. In ref. \cite{tak}
where this possibility is first pointed out there is no restriction on
parity given. Therefore in the special case $\gamma=\pi/N$ the 
meaning of the string hypothesis as taken from refs
\cite{tak} is that the form for the roots is (\ref{sform})
and the allowed string length are $(1,+),\dots ,(N,+),~(1,-),~(N,-).$
We note, however, that implicitly certain complete $N$ strings
are contained in the combinatorial completeness 
arguments of ref. \cite{kirlis} because this counting contains
 composite states consisting of an $(N-1,+)$ string and a $(1,-)$
 string and for $N\geq 3$ this composite state will in fact be a
 complete N string if the real parts of $(N-1,+)$ and $(1,-)$ are equal.

There is however, a very important feature of these 
complete $N$ strings which has not been clearly stated previously
because in equations (197) and (198) of ref. \cite{tak} only the limiting
$L\rightarrow \infty$ form of the strings is given. In fact it
will be seen below both by examining the functional equation of Baxter
(\ref{feq}) and the numerical solutions of sec. 3  that the spacings of the
$N$ different imaginary parts of the complete $N$ strings are exactly
given by $2\gamma$ for finite $L.$  This feature is not 
present for all of the other
strings and we will thus refer to these $N$ strings as ``exact
complete $N$ strings''. 

From (\ref{energy}) we see that the contribution of an exact complete N
string to the energy is zero (independent of $L$). 
In other words the eigenstates of the
Hamiltonian which differ only by exact complete N strings are
degenerate in energy. This is exactly the degeneracy which arises
from the $sl_2$ loop symmetry algebra found in ref.\cite{dfm}. From
(\ref{mom}) we find the contribution to the momentum of the exact
complete N string with $\gamma=\pi m/N$
\begin{equation}
P=\cases{0&if $N-m$ is even\cr
         \pi&if $N-m$ is odd}
\end{equation} 

The feature of the exact spacing of the imaginary
parts of the roots of the exact complete $N$ strings
has the dramatic effect that 
there are now terms in the Bethe's
equation (\ref{beq}) which are of the form $0/0$ and therefore there is
no equation left to determine the real part. Even worse there is no
equation to guarantee that $v_j^r$ need be real if there is a state
which contains two or more exact complete $N$ strings. 
We thus conclude that even though ref. \cite{tak} introduces exact
complete $N$ strings it actually computes nothing about them.    

\subsection{Exact complete $N$ strings in Baxter's $Q$}

The meaning and the necessity of exact complete $N$ strings comes from
the $sl_2$ loop algebra symmetry and is easily seen when Baxter's
matrix functional equation (\ref{feq})is written in terms of the eigenvalues of
$T(v)$ and $Q(v).$ When the root of unity condition $q^{2N}=1$ holds the
transfer matrix has degenerate eigenvalues. But on the other hand the
matrix $Q(v)$ does not have degenerate eigenvalues. Therefore 
the only way for the functional equation (\ref{feq}) 
to hold for the degenerate
eigenvalues of $T(v)$ with several distinct polynomials $Q(v)$ 
is for $Q(v)$ to contain factors of the exact complete $N$ string
\begin{equation}
Q_N(v)=\prod_{j=1}^N\sinh{1\over 2}(v-\alpha-2ji\gamma).
\label{ecns}
\end{equation} 
These factors obey $Q_N(v)=Q_N(v+2i\gamma)$ and hence 
the simultaneous vanishing condition
(\ref{vanishing}). Therefore we conclude  when the root of unity
condition (\ref{root}) holds that the
functional equation (\ref{feq}) is not sufficient to determine the
parameter $\alpha$ in the exact complete $N$ strings 
(\ref{ecns}) which exist if $L$ is sufficiently large.
Indeed the functional equation does not by itself even
guarantee that the imaginary part of $\alpha$ is either $0$ or $\pi.$

\subsection{Solution by Continuity}

For any fixed $L$ any deviation of $q$ from $q^{2N}=1$
will break all the degeneracies of the eigenvalues of $T(v)$ and now
there will be a one--to--one relation between the eigenvalues of $T(v)$
and $Q(v).$ Therefore one way to determine the values of $\alpha$ in
the exact complete $N$ string of (\ref{ecns}) is by continuity from
the  nondegenerate case. This will give a limiting set of
solutions of the Bethe's ansatz equation at the roots of unity.

In principle for any given root of unity we can analytically
determine a set of limiting Bethe's ansatz equations by continuity
from (\ref{beq}) which resolves the ambiguity of $0/0.$ However, here
we will follow an alternative numerical procedure which begins with the
functional equation (\ref{feq}).
Our procedure is as
follows. Because of the commutation relation $[T(v),T(v')]=0$ the eigenvectors
of $T(v)$ are independent of $v$ and we may determine them numerically
on the computer by choosing any convenient value of $v$ we please.
By letting $T(v)$ act on these 
$v$ independent numerical vectors we may determine the 
eigenvalues as polynomials in $e^{v}$ of degree $L$
on the computer. We then determine the coefficients of the ${L\over
2}-|S^z|$ degree polynomial for $Q(v)$ by considering the functional
equation (\ref{feq}) at ${L\over 2}-|S^z|+1$ different values of the
spectral parameter $v$ and solving the resulting system of homogeneous
linear equations on the computer. As long as the eigenvalues of $T(v)$
are non degenerate this procedure is unique and unambiguous. The
zeroes $v_j$ of the eigenvalues of $Q(v)$ are then easily found by
finding the roots of the ${L\over 2}-|S^z|$ order polynomials on the computer. 
The limit of $\Delta\rightarrow 0$ and $\Delta \rightarrow \pm 1/2$ is
then obtained by studying sequences of $\Delta$ which approach the root
of unity under consideration. Of course in practice there is an
optimum value of $\Delta$ such that if we come closer to the root of
unity than this value the accuracy of the computation will
deteriorate. Fortunately for the cases considered in this paper this
necessary limitation does not interfere with our ability to see the
qualitative features of the limiting case.

This method of continuity can be done for any value of $S^z.$ The
validity of this approach to the limit is then checked for the special
case $S^z=0$ by numerically diagonalizing the matrix (\ref{qform}) exactly
at $\Delta=0~(\gamma=\pi/2)$ and $\Delta=-{1\over 2}~(\gamma=\pi/3).$

\section{Numerical studies for $L=16$}

In this section we use the procedure outlined above to study the
continuous solution to Bethe's equation (\ref{beq}) at $\Delta=0$ and
$\Delta=-1/2.$ The two cases are presented separately.

\subsection{$\Delta=0~(\gamma=\pi/2)$}

We have obtained the Bethe's roots for all $2^{16}$ eigenvalues
of the $L=16$ chain from the $Q$ of
(\ref{qform}) at exactly $\Delta=0$ for $S^z=0$ and by the limiting
procedure described above for all other values of $S^z.$
The root content of all eigenvalues without exception is described by
$(1,+),~(1,-)$ strings, exact complete 2 strings and infinite roots.
The ground state  contains only $(1,+)$ strings and its root content
is given in table 1.

 The
states are grouped into degenerate multiplets which have a highest
weight (in terms of $S^z$) which  have only $(1,+)$ and $(1,-)$ roots
in the sector $S^z=0 ({\rm mod}~2)$ but which may also contain an
infinite root for $S^z=1 ({\rm mod}~2).$ 
The highest weight states are not degenerate and are therefore
identical with the states of ref.\cite{lsm}.

The remaining members of the multiplet
contain exact complete 2 strings. The content of exact complete 2
strings and large roots for each type of multiplet is given in table
2. 

In table 3 we give examples of multiplets with one exact complete two string.

In table 4 we consider multiplets with multiple exact complete two strings by 
considering the states
whose highest weight is $S^z=6$ 
 with momentum $2\pi/ 16$ in $S^z=0,\pm 4$ and $18\pi/16$ in $S^z=\pm
2,\pm 6$
There are seven such multiplets with
energies $0, \pm .2986..., \pm .72111...,\pm .55197....$ These
multiplets have 6 states in $S^z=\pm 4,$ 15 states in $S^z=\pm 2$ and
20 states in $S^z=0.$ In table 4  we give the root content of the state
with $E=-.55197$ in $S^z=0$ and $S^z=2$ as an illustration.
In particular we note that the following types of
imaginary parts occur

1) $0$ and $\pi$

2) $\pm \pi/2$

3) Pairs of two strings with imaginary parts other than $0,\pm \pi/2$
and $\pi.$

In figure 1 we extend this by plotting the location of all
Bethe's roots of all eigenvalues in the sector $S^z=0.$ In this plot
all roots whose imaginary part is not $0$ or $\pi$ come from exact
complete 2 strings.

In table 5 we give examples of multiplets with $S^z=3,1,-1$ and $S^z=3,1,-1,-3$

With the data from fig. 1 we may now examine  in much greater detail
the meaning of the term''Bethe's ansatz equation'' which is commonly
used to refer to eqn. (\ref{beq}). Consider first directly setting
$\gamma=\pi/2$ in (\ref{beq}). 
We see that (\ref{beq}) reduces to
\begin{equation}
\left({\sinh{1\over 2}(v_j+{\pi i\over 2})\over \sinh{1\over
2}(v_j-{\pi i\over 2})}\right)^L=(-1)^{{L\over 2}-|S^z|-1}  
\label{lsm1}
\end{equation}
and thus
\begin{equation}
{\sinh{1\over 2}(v_j+{\pi i\over 2})\over \sinh{1\over 2}(v_j-{\pi
i\over 2})}=\cases{e^{2\pi i m\over L}&if ${L\over 2}-|S^z| \equiv 1 ({\rm
mod}~2)$\cr
e^{{2\pi i\over L}(m+1/2)}& if ${L\over 2}-|S^z|\equiv 0 ({\rm mod}~2)$\cr}
\label{lsm2}
\end{equation}
and from (\ref{pdef})
\begin{equation}
p_{m} = 
\cases{{2\pi i m\over L}&if ${L\over 2}-|S^z| \equiv 1 ({\rm
mod}~2)$\cr
{2\pi i\over L}(m+1/2)& if ${L\over 2}-|S^z|\equiv 0 ({\rm mod}~2)$\cr}
\label{lsm3}
\end{equation}
with $m=0,1,\cdots, L-1.$ 
The solutions to (\ref{lsm2}) all satisfy
\begin{equation}
\Im v_j=0,~{\rm or}~\pi
\end{equation}
which obviously is in gross contradiction to the example of table 2
and with the large mass of data summarized in Fig. 1.
On the other hand the roots of (\ref{lsm1}) are precisely the roots of
the string ansatz of ref. \cite{kirlis} and agree exactly with the
computation of Lieb, Schultz and Mattis \cite{lsm}.

This disagreement between the solution of Bethe's equation (\ref{beq})
for $\Delta$ taken continuously to zero and the solutions of
(\ref{lsm1}) was noted in ref. \cite{nlk} where the equation
(\ref{lsm1})is referred to as Bethe's equation and with this terminology  
the authors are able to say that Bethe's equation is complete at
$\Delta=0$ but that there is a discontinuity at $\Delta=0.$
However, as we emphasized in sec. 1 the Bethe's equation (\ref{beq})
is only derivable under the assumption that the simultaneous
vanishing (\ref{vanishing}) does not occur. It seems to us more
appropriate to preserve the condition (\ref{vanishing}) whenever we
refer to (\ref{beq}) as the ``Bethe's equation.''
 Since the equation (\ref{lsm1}) holds even for the cases where the
simultaneous vanishing (\ref{vanishing}) occurs we would prefer to
call it the Lieb-Schultz-Mattis equation after is original discoverers.
With this terminology we reserve the name ``Bethe's equation'' at roots of
unity as the equation satisfied by the roots obtained by continuity as
$\Delta\rightarrow 0$ for $S^z\neq 0$ and by the roots of $Q(v)$ of
(\ref{qform}) at $\Delta$ exactly zero for $S^z=0.$ 
With his terminology Bethe's equation will be continuous at $\Delta=0$
by definition; however the explicit form of the equation when there are
exact complete 2 strings in the state is not known. 
Regardless of terminology it is a fact that the roots of the 
eigenvalues of $Q(v)$ given by
(\ref{qform}) are not all given by (\ref{lsm1}).

\subsection{$\Delta=-1/2~(\gamma=\pi/3)$}

If this lack of continuity happened only at $\Delta=0$ it would
perhaps only be of semantic interest whether or not we call (\ref{beq})
without the condition (\ref{vanishing}) by the name of Bethe's
equation. But the phenomenon which we just saw at $\Delta=0$ happens for all
$\Delta$ obtained from the root of unity condition (\ref{root}). To
make this specific we here explicitly consider the case $\Delta=-1/2.$

The root contents of the highest weight
state of each multiplet are now made up  of $(1,+), (2,+)$ and $(1,-)$
strings. 
In table 6 we give the root content of the
ground state for $\Delta=\pm 1/2$ which contains only 
$(1,+)$ roots and the excited state 
(in $P=4\pi/16$) which contains a single $(2,+)$ string.
In table 7 we list the exact complete 3 string and infinite
root content of all multiplets. In table 8 we give several examples
of multiplets with one exact complete 3 string and
in table 9 we give an example of
a multiplet with $S^z=6,3,0,-3,-6$
In table 10 we give examples of multiplets 
with $S^z=5,2,-1,-4,~~S^z=5,2,-1$ and $S^z=4,1,-2$

In Fig. 2  we plot the position of the roots of all eigenvalues of
$Q(v)$ in  the sector $S^z=0.$ We note as for $\Delta=0$ that the
values of the imaginary parts which are not $0,~\pm\pi/3$ or $\pi$ are
all for roots of exact complete 3 strings.

\section{Completeness, Incompleteness and evaluation parameters}

As done above for $\Delta=0$  it is always possible to make 
the statement that Bethe's
equation is complete in the degenerate cases if we can start from 
a completeness in the case
where the parameter $\Delta$ is generic and then define the term 
``Bethe's equation'' at the root of unity case by continuity. This, of
course, has the serious disadvantage that at roots of unity most of
the ``Bethe's equations'' are not yet known.

On the other hand if we define Bethe's equation to be (\ref{beq}) with
the restriction (\ref{vanishing}) even at roots of unity (\ref{root})
then the examples given above for $\Delta=-1/2$ demonstrate that
Bethe's equation by itself is not complete. Instead with this
definition Bethe's equation is complete for the highest weight of the
multiplet and the remaining states must be obtained by applying
appropriate lowering operators to this state of highest $S^z$.
 
To be more precise we turn to the theory of finite dimensional
representations of (quantum) affine Lie algebras \cite{char}-\cite{cp}
 where the states of a
degenerate multiplet are specified by tensor products of evaluation
representations. To determine the evaluation parameters of these
representations (in the sector $S^z\equiv 0~({\rm mod}~N))$
we use the two  Chevalley generators of the
$sl_2$ loop algebra $T^{+(N)}$ and $S^{-(N)}$ defined in
ref. \cite{dfm} to define the numbers $\mu_r$ from
\begin{equation}
{T^{+(N)}\over r!}{S^{-(N)}\over r!}\Omega=\mu_r\Omega
\end{equation}
where $\Omega$ is the vector in the multiplet with the maximum value
of $S^z.$ From these $\mu_r$ we form the Drinfeld polynomial
\begin{equation}
P(x)=\sum_{r\geq 0}\mu_r(-x)^r.
\end{equation}
Then the evaluation parameters $a_j$ are given as
\begin{equation}
P(x)=\prod_j(1-a_j x).
\end{equation}

As a specific example Jimbo (private communication) has shown that for
a chain with $L\equiv 0~({\rm mod}~N)$ with $N$ odd and 
$q=e^{\pi i/N}$ that for the multiplet whose highest weight state
is the vector with all spins up we have
\begin{equation}
\mu_r={L!\over (rN)!(L-rN)!}
\end{equation}
and thus the corresponding Drinfeld polynomial is
\begin{equation}
P(x)={1\over N}\sum_{k=0}^{N-1}(1-e^{2\pi i/N}x^{1/N})^L
\end{equation}

For a complete solution to the problem we need an efficient method of
computing the evaluation parameters and an explicit construction of
the eigenvectors of $T(v)$ in terms of these evaluations
parameters. Both of these problems are open at the present.

Moreover the relation of the evaluation representation to the exact
complete N strings is not known. We do know empirically that for the
multiplets $S^z=N,0,-N$ (which have two states in $S^z=0$ and
 are specified by two evaluation
parameters $a_1,~a_2$) that the real parts of the complete exact N
strings  of both the states in $S^z=0$ are equal and are given by
\begin{equation}
\alpha={1\over 2N}{\rm ln}a_1a_2.
\end{equation}
We also know for the multiplet $S^z=2N,N,0,-N,-2N$ 
that if we consider the states in $S^z=0$ with two complete exact $N$ strings
which have the same real part then  this real part is given in terms
of the four evaluation parameters as
\begin{equation}
\alpha={1\over 4N}{\rm ln}\alpha_1\alpha_2\alpha_3\alpha_4
\end{equation}

But for $\Delta=0$ it was possible to go further and
produce the alternative equation (\ref{lsm1}) to reduce (\ref{beq}) to
(\ref{lsm1}) and this equation is well defined even for solutions
where the simultaneous vanishing (\ref{vanishing}) occurs. 
This equation did not give the roots of $Q(v)$ but did correctly give
all eigenvalues and eigenvectors of the Hamiltonian. 
It can also be determined that all evaluation parameters are of the
form
\begin{equation}
a_j=\cot ^2{1\over 2}(p_j+{\pi\over 2})
\end{equation}
where
$p_j$ is given by (\ref{lsm3}).

It is thus natural to ask if a
similar procedure can be done for other values of $\Delta.$ 
This seems to be the approach to completeness taken in ref.\cite{kirlis}
where a string ansatz is made for the solutions of (\ref{beq})
whenever the equation is well defined and then (tacitly) applied for the cases
where the vanishing condition (\ref{vanishing}) holds and the
equations is not defined. This procedure gave the correct counting
but because no equation comparable to (\ref{lsm1}) is given there it is
 not possible to construct the corresponding degenerate states in the
multiplet. Furthermore the relation which this procedure has to 
the  evaluation representation of the $sl_2$ loop algebra symmetry is
not known.

\bigskip

 \centerline{{\bf Acknowledgments}}

 \bigskip
 This work is supported in part by the National Science Foundation
 under Grant No. DMR--0073058. We are most pleased to acknowledge
 useful discussions with R.J. Baxter,  M. Jimbo, A. Kl{\"u}mper, V. Korepin,
 Y. St. Aubin and  P. Zinn--Justin. We also wish to
 thank I.G. Korepanov for bringing references
 \cite{korep1},\cite{korep2} and \cite{korep3} to our attention.

\newpage

{\bf Tables}

Table 1. 

Root content for $\Delta=0$  of the ground state  with $ E =  -10.251\cdots$
 and $ P=0$ 

\begin{tabular}{|r|}\hline
$ -2.317786$\\
$ -1.192878$\\
$ -0.626402$\\
$ -0.197623$\\
$  0.197623$\\
$  0.626402$\\
$  1.192878$\\
$  2.317786$\\ \hline
\end{tabular}
\newpage

{\footnotesize
Table 2. Types of degeneracies of the transfer matrix $T(u)$ for the
case $\Delta=0$ and $L=16.$ 

\begin{tabular}{|rrl|}\hline
\multicolumn{3}{|l|}{Maximum $S^z=8$, one type}\\ \hline
$S^z= 8$&   multiplicity=        1& \\
$ 6$&           8&   one 2 string\\
$ 4$&           28&   two 2 strings\\
$2$&          56&   three 2 string\\
$0$&          70&   four 2 strings\\ 
$-2$&          56&   three 2 strings\\ 
$-4$&          28&   two 2 strings\\ 
$-6$&          8& one 2 string  \\ 
$-8$&          1&   \\ \hline \hline
\multicolumn{3}{|l|}{Maximum $S^z=7$, one type}\\ \hline
$S^z= 7$&     multiplicity=      1&  \\
$ 5$&           7&  one 2 string\\
$ 3$&           21&  two 2 strings\\
$1$&           35&  three 2 strings \\
$-1$&           21&  two infinite roots, two 2 strings\\ 
$-3$&           7& 2 infinite roots, one 2 string\\ 
$-5$&           1& 2 infinite roots \\ \hline \hline
\multicolumn{3}{|l|}{Maximum $S^z=6$, one type}\\ \hline
$S^z= 6$& multiplicity=           1&\\
$ 4$&            6& one 2 string\\
$ 2$&            15& two 2 strings\\
$ 0$&            20& three 2 strings\\
$-2$&            15& two 2 strings\\ 
$-4$&            6&one 2 string\\ 
$-6$&            1&\\ \hline \hline
\multicolumn{3}{|l|}{Maximum $S^z=5$ two types}\\ \hline
Type 1&&\\ 
$S^z= 5$& multiplicity=          1&\\
$ 3$&           4& one 2 string\\
$1$&           6& two 2 strings\\ 
$-1$&           4& two infinite roots, 2 strings\\ 
$-3$&           1& two infinite roots, one 2 string\\ \hline 
Type 2&&\\
$S^z= 5$& multiplicity=          1&one infinite root\\
$ 3$&           5& one infinite root, one 2 string\\
$1$&           10& one infinite roots, two 2 strings\\ 
$-1$&           10& one infinite roots, two 2 strings\\ 
$-3$&           5& one infinite roots, one 2 string\\ 
$-5$&           1& one infinite root\\ \hline \hline
\end{tabular}

\newpage

Table 2 continued

\begin{tabular}{|rrl|}\hline
\multicolumn{3}{|l|}{Maximum $S^z=4$ one type}\\ \hline
$S^z= 4$& multiplicity=           1&\\
$ 2$&            4& one 2 string\\ 
$0$&            6& two 2 strings\\ 
$-2$&            4& one 2 string\\ 
$-4$&            1& \\ \hline \hline
\multicolumn{3}{|l|}{Maximum $S^z=3$, two types}\\ \hline
$S^z= 3$&  multiplicity =          1&  \\
$ 1$&            2& one 2 string\\
$-1$&          1& two infinite roots  \\ \hline 
$S^z= 3$&  multiplicity =          1&one infinite root  \\
$ 1$&            3&one infinite root, one 2 string\\
$-1$&          3& one infinite root, one 2 string  \\ 
$-3$&            1&one infinite root\\ \hline \hline
\multicolumn{3}{|l|}{Maximum $S^z=2$}\\ \hline
$S^z= 2$& multiplicity=          1& \\
$0$&           2& one 2 string \\ 
$-2$&           1&\\ \hline \hline
\multicolumn{3}{|l|}{Maximum $S^z=1$ two types}\\ \hline
$S^z= 1$& multiplicity=          1& (non degenerate)\\ \hline 
$S^z= 1$& multiplicity=          1&one infinite root \\ 
$S^z= -1$& multiplicity=          1&one infinite root \\ \hline
\multicolumn{3}{|l|}{Maximum $S^z=0$ one type}\\ \hline
$S^z= 0$ & multiplicity=         1& (nondegenerate)\\ \hline \hline
\end{tabular}}

\newpage
{\footnotesize
Table 3. Examples of multiplets with $S^z=2,0,-2$ in $L=16$ and
$\Delta=0$ with one exact complete 2 string

$E=-8.750\cdots$

\begin{tabular}{|rl|rl|} \hline  
\multicolumn{4}{|l|}{$S^z=2,~P=18\pi/16$}\\ \hline      
$ - 1.1927968$&&&\\
$ - 0.6263626$&&& \\
$ - 0.1976133$&&&\\
$   0.1976060$&&&\\
$   0.6263537$&&&\\
$   2.3175502$&&&\\ \hline
\multicolumn{4}{|l|}{$S^z=0,~P=2\pi/16$}\\ \hline                     
$ -1.192878$&&$-1.1928789$&\\   
$-0.626402$&&$-0.6264027$&\\ 
$ -0.197623$&&$-0.1976234$&\\
$  0.197623$&&$ 0.1976234$&\\
$  0.626402$&&$ 0.6264027$&\\
$  2.317786$&&$ 2.3177860$&\\
$ -0.562453$&$+i\pi/2$&$-0.562453$&\\
$ -0.562453$&$ -i\pi/2$&$-0.562453$&$ +i\pi$\\ \hline 
\end{tabular}

\vspace{.2in}

$E= -7.249\cdots$

\begin{tabular}{|rl|rl|} \hline  
\multicolumn{4}{|l|}{$S^z=\pm 2,~P=18\pi/16$}\\ \hline  
$-  2.317555$&&&\\
$  -0.626359$&&&\\
$  -0.197613$&&&\\
$   0.197602$&&&\\
$   0.626344$&&&\\
$   2.317902$&$+i\pi$ &&\\ \hline
\multicolumn{4}{|l|}{$S^z=0,~P=2\pi/16$}\\ \hline
$-2.317786 $&    &$ -2.317786$&\\
$-0.626402 $&    &$ -0.626402$&\\
$-0.197623$ &    &$ -0.197623$&\\
$ 0.197623$ &     &$ 0.197623$&\\
$ 0.626402$ &     &$ 0.626402$&\\
$ 2.317786$&$ +i\pi$&$ 2.317786$&$ +i\pi$\\ 
$ 0.0 $&&$0.0$&$+i\pi/2  $   \\
$ 0.0$&$ +i\pi$& $0.0$&$ -i\pi/2 $     \\ \hline
\end{tabular}}
\newpage

{\footnotesize Table 4.

An example of the roots of a degenerate multiplet $S^z=6,4,2,0,-2,-4,-6$ 
in $\Delta=0,~L=16$ with highest
weight $S^z=6.$ We choose the state with  $E=-.5517987.....$ 
The multiplet has $2^6=64$ states. In $S^z=0$  there are 20 states,
 the momentum is $P=2\pi/16$
and  the roots 
are computed from the explicit expression for
$Q$ (\ref{qform}). 
In $S^z=2$ there are 15 states, the momentum is $P=18\pi/16.$
All states have the same (1,+) and (1,-)
roots in addition to  exact complete 2 strings.

\begin{tabular}{|rl|rl|}\hline
\multicolumn{4}{|l|}{$S^z=0,~P=2\pi/16$}\\ \hline 
\multicolumn{2}{|l|}{$\sum \Im v_j\equiv 0~({\rm mod}~2\pi)$}
&\multicolumn{2}{l|}{$\sum \Im v_j\equiv
\pi~({\rm mod}~2\pi)$}\\ \hline
$  0.626402      $&&    $0.626402$&   \\  
 $ 1.192878$&$+i\pi   $ &$ 1.192878 $&$+i\pi$\\  
$ -1.994484$&$+i\pi/2 $ &$-2.006672$&$+i\pi/2$\\
$ -1.994484$&$-i\pi/2 $&$ -2.006672$&$-i\pi/2 $\\
$ -0.354311$&$+i\pi/2$&$  -0.409599$&$ +i\pi/2$ \\
$ -0.354311$&$-i\pi/2$&$  -0.409599$&$ -i\pi/2$ \\
$  1.439155   $&&$       1.506631 $&$+i\pi/2$\\
$  1.439155$&$+i \pi$&$    1.506631$&$ -i\pi/2$\\\hline
 $ 0.626402$&&$          0.626402$&\\
 $ 1.192878$&$ +i\pi$  &$  1.192878$&$ +i\pi$\\
 $-1.877737$&$ +i\pi/2$&$ -1.862425$&$+i\pi/2$  \\
 $-1.877737$&$ -i\pi/2$&$ -1.862425$&$-i\pi/2 $ \\
 $ 0.044698$&$ +i\pi$&$    0.067738 $ & \\
 $ 0.044698 $ &&$        0.067738 $&$+i\pi$\\
 $ 0.923398 $&$+i\pi/2$& $ 0.885045 $& \\
 $ 0.923398 $&$-i\pi/2$ &$ 0.885045 $&$+i\pi$  \\ \hline
$  0.626402 $&&$         0.626402 $  & \\
$  1.192878 $&$+i\pi  $&$  1.192878$&$+i\pi  $\\
$ -1.796958 $&$+i\pi/2$&$ -1.788035$&$+i\pi/2$ \\
$ -1.796958 $&$-i\pi/2$&$ -1.788035$&$-i\pi/2 $\\
$ -0.415405 $&$+i\pi $&$  -0.409732$&\\
$ -0.415405  $&&$       -0.409732$&$+i\pi$\\
$  1.302723 $&$+i\pi/2$& $ 1.288127$&\\
$  1.302723 $&$ -i\pi/2$& $1.288127$&$+i\pi$\\ \hline
$  0.626402$ &&$         0.626402$  &\\
$  1.192878$&$+i\pi$&$     1.192878$&$+i\pi$ \\ 
$ -1.852568$&$ +i\pi$& $  -1.847467$&\\
$ -1.852568  $&& $      -1.847467$&$+i\pi$\\
$ -0.538335 $&$+i\pi/2$& $-0.482661$&$+i\pi/2$\\ 
$ -0.538335 $&$ -i\pi/2$&$-0.482661$&$+i\pi/2$\\
$  1.481262 $&$+i\pi/2 $& $1.420488$&\\
$  1.481262 $&$-i\pi/2$ &$ 1.420488$&$+i\pi$\\ \hline
\end{tabular}

\newpage
Table 4 continued

\begin{tabular}{|rl|rl|}\hline
\multicolumn{4}{|l|}{$S^z=0,~P=2\pi/16$}\\ \hline 
\multicolumn{2}{|l|}{$\sum \Im v_j\equiv 0~({\rm mod}~2\pi)$}
&\multicolumn{2}{l|}{$\sum \Im v_j\equiv
\pi~({\rm mod}~2\pi)$}\\ \hline
$  0.626402$    &   &$0.626402$&   \\
$  1.192878 $&$+i\pi$ &$1.192878$&$+i\pi$\\
$ -1.765722  $&&$    -1.768190$&\\
$ -1.765722 $&$+i\pi$&$-1.768190$&$+i\pi$\\ 
$ -0.421083  $&&$    -0.426754$  & \\
$ -0.421083$&$+i\pi$&$ -0.426754$&$+i\pi$   \\ 
$  1.277164  $&& $    1.285303$&$+i\pi/2$  \\
$  1.277165 $&$+i\pi$&$ 1.285303$&$-i\pi/2$  \\ \hline
$  0.626402$  && $    0.626402$&\\
$  1.192878 $&$+i\pi$& $1.192878$&$+i\pi$\\
$ -1.793415 $&&$     -1.799117$&\\
$ -1.793415 $&$+i\pi$&$-1.799117$&$+i\pi$\\
$  0.067785  $&& $    0.040394$&\\
$  0.067785 $&$+i\pi$& $0.040394$&$+i\pi$\\ 
$  0.815989  $&&$     0.849081$&$+i\pi/2$\\
$  0.815989$&$+i\pi$&$  0.849081$&$-i\pi/2$\\ \hline
$  0.626402   $&&$    0.626402$&\\
$  1.192878 $&$+i\pi$& $  1.192878$&$+i\pi$\\ 
$ -1.342393$&$ +i\pi/2$&$-1.307213$&$+i\pi/2$\\
$ -1.342393 $&$-i\pi/2$&$-1.307213$&$-i\pi/2$\\
$ -0.964462  $&&      $-0.960063$&\\
$ -0.964462$&$+i\pi$&   $-0.960063$&$+i\pi$\\
$  1.397214$&$ +i\pi/2$&$ 1.357636$&\\
$  1.397214$&$ -i\pi/2$&$ 1.357636$&$+i\pi$\\ \hline
\end{tabular}

\newpage
Table 4 continued

\begin{tabular}{|rl|rl|}\hline
\multicolumn{4}{|l|}{$S^z=0,~P=2\pi/16$}\\ \hline 
\multicolumn{2}{|l|}{$\sum \Im v_j\equiv 0~({\rm mod}~2\pi)$}
&\multicolumn{2}{l|}{$\sum \Im v_j\equiv
\pi~({\rm mod}~2\pi)$}\\ \hline
$  0.626402$ &               & $ 0.626402$&   \\
$  1.192878 $&$+i\pi$          & $ 1.192878$&$  +i\pi$\\ 
$ -1.122273$&$+ 0.163271i$&$ -1.137586$&$+   0.146097 i$\\
$ -1.122273$&$ -2.978321 i$ & $-1.137586$&$ -2.995495 i$\\
$ -1.122273$&$ -0.163271i$  &$ -1.137586$&$ -0.146097 i$\\
$ -1.122273$&$+ 2.978321i$& $-1.137586  $&$ +2.995495 i$\\
$  1.334906 $  &               &$ 1.365531$&$  +i\pi/2$\\
$  1.334906$&$ +i\pi$            &$ 1.365531 $&$ -i\pi/2$\\ \hline
$  0.626402   $&               &$  .626402 $&\\
$  1.192878$&$+i\pi$             &$ 1.192878$&$ +i\pi$\\  
$ -0.384740  $&              &$  -0.152129 $&$ +i\pi/2$\\
$ -0.384740$&$ +i\pi$           &$ -0.152129  $&$-i\pi/2$\\  
$ -0.262449 $&$ -2.561390 i$  &$ -0.378755$&$ -2.967736 i$\\
$ -0.262449 $&$ -0.580202 i$&$ -0.378755   $&$ -.173856 i$\\
$ -0.262449 $&$ +2.561390 i$&$ -0.378755 $&$ + 2.967736 i$\\
$ -0.262449$&$ + 0.580202 i$  &$ -0.378755 $&$+0.173856i$\\ 
\hline
$  0.626402   $&   &$  0.626402 $&\\
$  1.192878 $&$+i\pi$& $ 1.192878$&$ +i\pi$\\ 
$ -0.805653 $  &   &$ -0.857340$  &\\
$ -0.805653 $&$+i\pi$               &$ -0.857340$&$+i\pi$  \\
$ -0.051993  $&$ -2.957473 i$&$  0.000361  $&\\
$ -0.051993$&$+   0.184119 i$ &$ -0.052661 $&$ +i\pi/2$\\
$ -0.051993$&$ +  2.957473 i$  &$  0.000361$&$+i\pi  $\\
$ -0.051993 $&$  -0.184119 i$  &$ -0.052661 $&$ -i\pi/2$\\ \hline
\end{tabular}

\newpage
Table 4 continued

\begin{tabular}{|rl|rl|}\hline
\multicolumn{4}{|l|}{$S^z=2,~P=18\pi/16$}\\ \hline 
\multicolumn{2}{|l|}{$\sum \Im v_j\equiv 0~({\rm mod}~2\pi)$}
&\multicolumn{2}{l|}{$\sum \Im v_j\equiv
\pi~({\rm mod}~2\pi)$}\\ \hline
   $0.626343$&&        $ 0.626355$&\\  
   $1.192856$&$ +i\pi$&    $ 1.192850$&$ +i\pi$\\
 $- 0.458130 $&$+i\pi/2$&$ - 0.447383 $&$ + 2.958193i$\\
 $- 0.458130 $&$-i\pi/2$&$ - 0.447383$&$   -2.958193i$\\
 $- 0.419474$&&$ - 0.447366$&$  -0.183219i$\\  
 $- 0.419450$&$ +i\pi$&$ - 0.447366 $&$+0.183219i$\\ \hline
 $  0.626329   $&&$ 0.626368$&  \\
  $ 1.192864 $&$ +i\pi$&$ 1.192846$&$ +i\pi$\\
  $ 0.046111 $&$+i\pi/2 $&$ 0.004442  $&\\
  $ 0.046111 $&$-i\pi/2 $&$ 0.004435 $&$+i\pi$\\
$ - 0.954798 $&&$ - 0.909222   $&\\
$ - 0.954712$&$ +i\pi$&$ - 0.909190$&$+i\pi$\\ \hline
$0.626364 $&&&\\ 
$ 1.192848$&$  +i\pi$&&\\
$ 0.028744$&&&\\  
$ 0.028697 $&$+i\pi$&&\\ \
$-  0.883297$&$ +i\pi/2$&&\\ 
$-  0.883297$&$ -i\pi/2 $&&\\ \hline
$   0.626275$&&$ 0.626279   $&\\   
$   1.192971$&$ +i\pi $&$ 1.192969$&$+i\pi$\\
$  -1.066835$&$ +i\pi/2 $&$  -0.937278  $&\\
$  -1.066835$&$ -i \pi/2$&$  -0.937214$&$  +i\pi$\\
$   1.983971 $&&$ 1.981627$&\\
$   1.984254 $&$+i\pi$&$ 1.981907$&$ +i\pi$ \\ \hline 
&&$0.626286   $&\\
&& $1.192966$&$+i\pi$\\
&&$- 0.375322  $&\\
&&$ -0.375280$&$ +i\pi$\\
&&$  1.978775  $&\\
&&  $1.979051$&$ +i\pi$\\ \hline
&&$ 0.626307  $&\\
&&$ 1.192960$&$ +i\pi$\\
&&$ 0.080561   $&\\
&&$ 0.080711$&$  +i\pi$\\
&&$ 1.972039  $&\\
&&$ 1.972307$&$  +i\pi$\\  \hline
\end{tabular}
\newpage

Table 4 concluded

\begin{tabular}{|rl|rl|}\hline
\multicolumn{4}{|l|}{$S^z=2,~P=18\pi/16$}\\ \hline 
\multicolumn{2}{|l|}{$\sum \Im v_j\equiv 0~({\rm mod}~2\pi)$}
&\multicolumn{2}{l|}{$\sum \Im v_j\equiv
\pi~({\rm mod}~2\pi)$}\\ \hline
$ 0.626323  $&&&   \\
$ 1.192867 $&$ +i\pi $&& \\
$ 0.198855 $&$+i\pi/2$&&  \\
$ 0.198855 $&$-i\pi/2$&&  \\
$ -2.040445   $&&& \\
$ -2.040077 $&$+i\pi$&& \\ \hline
&&$ 0.626277$&   \\
&&$ 1.192970 $&$+i\pi$ \\
&& $1.982814$ & \\
&& $1.983095$&$ +i\pi$ \\
&&$-2.037150$&\\
&&$-2.036788$&$+i\pi$\\ \hline
&&$ 0.626349$ & \\
&& $1.192853$&$ +i\pi$\\
&&$-0.462058   $&\\
&&$-0.461966$&$ +i\pi$\\
&&$-2.023406  $&\\
&&$-2.023066$&$ +i\pi$\\ \hline
&&$ 0.626367  $&\\
&&$ 1.192847 $&$+i\pi$\\
&&$ 0.017464  $&\\
&&$ 0.017436$&$ +i\pi$\\
&&$-2.032311  $&\\
&&$-2.031957$&$ +i\pi$\\ \hline
&&$ 0.626319  $&\\
&&$ 1.192870 $&$+i\pi$\\
&&$ 0.304376$&$ +i\pi/2$\\ 
&&$ 0.304376$&$ -i\pi/2$\\
&&$-1.151308$&$+i\pi/2$\\ 
&&$-1.151308$&$ -i\pi/2$\\ \hline
&&$ 0.626344  $&\\
&&$ 1.192855 $&$ +i\pi$\\
&&$-1.061563   $&\\
&&$-1.061281 $&$+i\pi$\\ 
&&$-1.970389  $&\\
&&$-1.970124  $&$+i\pi$\\ \hline 
\end{tabular}}
\newpage

{\footnotesize
Table 5a.

An example of a multiplet     
for $\Delta=0$ and $L=16$ with $S^z=3,1,-1$

\vspace{.1in}

$E=  -7.875\cdots$

\begin{tabular}{|rl|rl|}\hline
\multicolumn{4}{|l|}{$S^z=3,~P=2\pi/16$}\\ \hline
$  0.881310$&&&\\
$  0.403173$&&&\\
$  0.0  $&&&\\
$ -0.403166$&&&\\
 $-1.614741$&&& \\ \hline
\multicolumn{4}{|l|}{$S^z=1,~P=18\pi/16$}\\ \hline
$  0.881326$ &&$  0.881384  $&\\
$  0.403174 $ && $0.403901$&\\
$  0.0   $&&$0.0$&\\
$- 0.403184$&&$ - 0.403208$&\\  
$- 1.614809 $&&$      - 1.614809$&\\
$  0.366737$&$ +i\pi/2$&$ 0.366743$&\\  
$  0.366737$&$ -i\pi/2$&$ 0.366038$&$ +i\pi$\\ \hline
\multicolumn{4}{|l|}{$S^z=-1,~P=2\pi/16$}\\ \hline
$  0.881310$&&&\\
$  0.403173$&&&\\
$  0.0$&&&\\
$ -0.403166$&&&\\
$ -1.614741$&&&\\
$\infty$&&&\\
$-\infty$&&&\\ \hline
\end{tabular}

\vspace{.4in}

Table 5b.

An example of a multiplet     
for $\Delta=0$ and $L=16$ with $S^z=3,1,-1,-3$

\vspace{.1in}

$E= -5.875\cdots$

\begin{tabular}{|rl|rl|rl|}\hline
\multicolumn{6}{|l|}{$S^z=\pm 3,~P=2 \pi/16$}\\ \hline
$ 1.614739 $&&&&&\\  
$ 0.881307  $&&&&&\\
$ 0.403186$&&&&& \\
$-0.403383$&&&&& \\
$\infty$&&&&&\\ \hline
\multicolumn{6}{|l|}{$S^z=\pm 1,~P=18 \pi/16$}\\ \hline
$ 1.614739 $&  &$ 1.614736 $&&$ 1.614735  $&\\
$ 0.881307  $&&$  0.881300$&&$  0.881298  $&\\
$ 0.403186 $ &&$  0.403168$&&$  0.403164  $&\\
$-0.403383$ &&$   -0.403156$&&$ -0.403175$ &\\ 
$\infty$&&$\infty$&&$\infty$&\\
$-0.258692$&&$-1.260395$&&$      -1.323623$ &$+i\pi/2 $\\
$-0.258590$&$+i\pi$&     $-1.260335$&$+i\pi$& $-1.323623$&$ -i\pi/2$ \\ \hline
\end{tabular}

\newpage

{\footnotesize
Table 6. 

The root content for the ground state and the state with  one
$(2,+)$ string for $\Delta=\pm1/2$ for $L=16$ computed from $Q(v).$

$\Delta=-1/2~(\gamma=\pi/3)$

\begin{tabular}{|r|l|rl|} \cline{1-1} \cline{3-4}
$P=0$&&\multicolumn{2}{|l|}{$P=4\pi/16$}\\
$E=-12.08552\cdots$&&\multicolumn{2}{|l|}{$E=-10.137668\cdots$}\\
 \cline{1-1} \cline{3-4}
$   -1.563663$&&$-1.225672 $&$  -1.048691i$\\
$   -0.798288$&&$-1.225672$&$ +  1.048691i$\\
$   -0.418617$&&$-0.401731$&\\
$   -0.132015$&&$-0.117889$&\\
$    0.132015$&&$ 0.144763$&\\
$    0.418617$&&$ 0.431224$&\\
$    0.798288$&&$ 0.812236$&\\
$    1.563663$&&$ 1.582741 $&\\ \cline{1-1} \cline{3-4}
\end{tabular}

\vspace{.2in}
$\Delta=+1/2~(\gamma=2\pi/3)$

\begin{tabular}{|r|l|rl|} \cline{1-1} \cline{3-4}
$P=0$&&\multicolumn{2}{|l|}{$P=4\pi/16$}\\
$E=-8.8272\cdots$&&\multicolumn{2}{|l|}{$E=-7.8380\cdots$}\\
 \cline{1-1} \cline{3-4}
$ -3.020233 $&&$-2.986276$&\\   
$ -1.583084 $&&$-1.546300$&\\
$ -0.833007$&& $-0.789520$&\\  
$ -0.262945$&& $-0.205871$&\\  
$  0.262945$&& $ 0.349177$& \\  
$  0.833007$&& $ 0.996326$&\\ 
$  1.583084$&& $ 2.091232$&$+ 2.083219i$\\ 
$  3.020233$&& $ 2.091232$&$ -2.083219i$\\ 
\cline{1-1} \cline{3-4}
\end{tabular}}

\newpage

{\footnotesize 
Table 7. Types of degeneracies of the transfer matrix $T(u)$ and the
Hamiltonian  for  $\Delta=-1/2 ~(\gamma=\pi/3)$ and $L=16.$ 

\begin{tabular}{|rrl|}\hline
\multicolumn{3}{|l|}{Maximum $S^z=8$, one type}\\
$S^z= 8$&   multiplicity=        1& \\
$ 5$&           4&   one 3 string\\
$ 2$&           6&   two 3 strings\\
$-1$&           4&   one 3 string, 4 infinite roots\\
$-4$&           1&   four infinite roots\\ \hline \hline
\multicolumn{3}{|l|}{Maximum $S^z=7$, one type}\\
$S^z= 7$&     multiplicity=      1&  \\
$ 4$&           4&  one 3 string\\
$ 1$&           6&  two 3 strings\\
$-2$&           4&  one 3 string, two infinite roots\\
$-5$&           1&  two infinite roots\\ \hline \hline
\multicolumn{3}{|l|}{Maximum $S^z=6$, one type}\\
$S^z= 6$& multiplicity=           1&\\
$ 3$&            4& one 3 string\\
$ 0$&            6& two 3 strings\\
$-3$&            4& one 3 string\\
$-6$&            1&\\ \hline \hline
\multicolumn{3}{|l|}{Maximum $S^z=5$, two types}\\ 
\multicolumn{3}{|l|}{Type 1}\\
$S^z= 5$& multiplicity=          1&\\
$ 2$&           2& one 3 string\\
$-1$&           1& four infinite roots\\ \hline 
\multicolumn{3}{|l|}{Type 2}\\
$S^z= 5$&  multiplicity=         1& one infinite root\\
$ 2$&           3& one infinite root, one 3 string\\
$-1$&           3& two infinite roots, one 3 string\\
$-4$&           1& two infinite roots\\ \hline \hline
\multicolumn{3}{|l|}{Maximum $S^z=4$, one type}\\
$S^z= 4$& multiplicity=           1&\\
$ 1$&            2& one 3 string\\ 
$-2$&            1& two infinite roots\\ \hline
\end{tabular}

\newpage

Table 7 concluded

\begin{tabular}{|rrl|}\hline
\multicolumn{3}{|l|}{Maximum $S^z=3$, one type}\\
$S^z= 3$&  multiplicity =          1&  \\
$ 0$&            2& one 3 string\\
$-3$&            1&\\ \hline \hline
\multicolumn{3}{|l|}{Maximum $S^z=2$, two types}\\
\multicolumn{3}{|l|}{Type 1}\\ 
$S^z= 2$& multiplicity=          1& (nondegenerate)\\ \hline
\multicolumn{3}{|l|}{Type 2}\\
$S^z= 2$&  multiplicity=         1& one infinite root\\         
$-1$&           1& two infinite roots\\ \hline \hline
\multicolumn{3}{|l|}{Maximum $S^z=1$ one type}\\
$S^z= 1$& multiplicity=          1& (nondegenerate)\\ \hline\hline
\multicolumn{3}{|l|}{Maximum $S^z=0$ one type}\\
$S^z= 0$ & multiplicity=         1& (nondegenerate)\\ \hline
\end{tabular}}
\newpage

{\footnotesize
Table 8.

Examples of states with one exact complete 3 string
for $\Delta=-1/2~(\gamma=\pi/3)$

$E=  -8.3926\cdots$

\begin{tabular}{|rl|rl|rl|}\hline
&&\multicolumn{4}{|l|}{$S^z=0,~P=2\pi/16$}\\ \cline{3-6}
\multicolumn{2}{|l|}{$S^z=3,~P=2\pi/16$}&
\multicolumn{2}{l|}{$\sum \Im v_j\equiv 0~({\rm mod}~2\pi)$}& 
\multicolumn{2}{l|}{$\sum \Im v_j\equiv \pi~({\rm mod}~2\pi)$}\\ \hline
$  0.562469$&&$ 0.563138$&  &$  0.563138$ & \\
$  0.263924$&&$ 0.264228 $& &$  0.264228$ & \\
$  0.007100$&&$ 0.007099$&&$    0.007099$&\\
$ -0.248840$&&$-0.249147$ & &$ -0.249147$ & \\ 
$ -0.959569$&&$-0.960808$&&$   -0.960808$&  \\ 
&&$0.125163 $ &&$  0.125163$&$+i\pi$\\
&&$0.125163$&$ +i2\pi/3$ & $ 0.125163$&$+i\pi/3$\\
&&$0.125163$&$ -i2\pi/3$&$   0.125163$&$-i\pi/3 $\\ \hline
\end{tabular}

\vspace{.2in}

 $E= -3.1921\cdots$

\begin{tabular}{|rl|rl|rl|}\hline
&&\multicolumn{4}{|l|}{$S^z=0,~P=2\pi/16$}\\ \cline{3-6}
\multicolumn{2}{|l|}{$S^z=3,P=2\pi/16$}&
\multicolumn{2}{l|}{$\sum \Im v_j\equiv 0~({\rm mod}~2\pi)$}& 
\multicolumn{2}{l|}{$\sum \Im v_j\equiv \pi~({\rm mod}~2\pi)$}\\ \hline
$  1.795639$&& $ 1.798356   $&&$  1.798356   $& \\
$  0.570485$&&$  0.571105  $&&$   0.571105  $&\\
$  0.288133$&&$  0.288419  $&&$   0.288419  $&\\
$ -0.434281$&& $-0.434765  $&&$  -0.434765  $&\\
$ -0.517966 $&$+i\pi$&$-0.518450$&$ +i\pi$&$-0.518450$&$+i\pi$ \\
&&$ -0.568222$&$+i\pi$ &$ -0.568222 $&  \\
&&$ -0.568222$&$ +i\pi/3$&$ -0.568222$&$+i2\pi/3$ \\
&&$ -0.568222$&$ -i\pi/3$&$ -0.568222$&$-i2\pi/3$ \\ \hline
\end{tabular}

\vspace{.2in}
$E=-2.6958\cdots$

\begin{tabular}{|rl|rl|rl|}\hline
&&\multicolumn{4}{|l|}{$S^z=0,~P=2\pi/16$}\\ \cline{3-6}
\multicolumn{2}{|l|}{$S^z=3,P=2\pi/16$}&
\multicolumn{2}{l|}{$\sum \Im v_j\equiv 0~({\rm mod}~2\pi)$}& 
\multicolumn{2}{l|}{$\sum \Im v_j\equiv \pi~({\rm mod}~2\pi)$}\\ \hline
$  1.760626$& &$ 1.763480 $&&  $ 1.763480$&  \\
$  0.528642$&&$  0.529315$ && $  0.529315 $& \\
$  0.238033$&&$  0.238347$ && $  0.238347 $ & \\
$ -0.481042  $&$+i\pi/3$& $ -0.481597$&$ + i\pi/3$&$ -0.481597 $&$ + i\pi/3$\\
$ -0.481042$&$   -i\pi/3$&$ -0.481597$&$  -i\pi/3$&$-0.481597$&$  -i\pi/3$\\
&&$-0.522649$  && $-0.522649$&$+i\pi$ \\
&&$-0.522649 $&$+i2\pi/3$ & $-0.522649 $&$+i\pi/3$ \\
&&$-0.522649$&$ -i2\pi/3 $& $-0.522649$&$ -i\pi/3 $\\ \hline
\end{tabular}}

\newpage

{\footnotesize
Table 9 

A multiplet in $P=2 \pi/16$ with $S^z=6,3,0,-3,-6$
for $L=16$ and $\Delta=-1/2~(\gamma=\pi/3).$
The roots for $S^z=6$ and $3$ are taken for $\Delta=-.501.$ The roots
for $S^z=0$ are obtained from $Q(v)$ of (\ref{qform}) at exactly $\Delta=-1/2.$
The root content of the highest weight is $(1+),(1-)$ 
The energy is $E=  3.7394\cdots.$
\vspace{.2in}

\begin{tabular}{|rl|rl|} \hline
\multicolumn{4}{|l|}{$S^z=\pm 6$} \\ \hline
$  1.957258  $&&&\\
 $-2.540234$&$+i\pi$&&\\ \hline
\multicolumn{4}{|l|}{$S^z=\pm3$} \\ \hline
\multicolumn{2}{|l|}{$ \sum \Im v_j \equiv 0~({\rm mod}~2\pi)$}
&\multicolumn{2}{l|}{$\sum \Im v_j\equiv \pi~({\rm mod}~2\pi)$} \\ \hline
$  1.958725  $&&$      1.958567$&\\
$ -2.536625$&$ +i\pi$&$ -2.536538$&$+i\pi$\\ 
$  0.051352$&$ +i\pi$&$  0.058681$&$+i 2\pi/3$\\ 
$  0.051227$&$ +i\pi/3$&$0.058681$&$-i 2\pi/3$\\
$  0.051227 $&$-i\pi/3$&$0.058681$& \\  \hline
$  1.958683$&&&\\
$ -2.536888$&$ +i\pi$&& \\
$ -0.898706$&$ +i \pi$&&\\ 
$ -0.894563$&$ +i \pi/3$&&\\
$ -0.894563$&$ -i\pi/3 $&&\\ \hline
$  1.959915  $&&&\\
$ -2.536608$&$+i\pi$&&\\ 
$  1.171936$&$ +i\pi$&&\\
$  1.165390$&$ +i\pi/3$&&\\
 $ 1.165390$&$ -i\pi/3$&&\\ \hline
\end{tabular}

\newpage

Table9 concluded

\begin{tabular}{|rl|rl|} \hline
\multicolumn{4}{|l|}{$S^z=0$} \\ \hline
\multicolumn{2}{|l|}{$ \sum \Im v_j \equiv 0~({\rm mod}~2\pi)$}
&\multicolumn{2}{l|}{$\sum \Im v_j\equiv \pi~({\rm mod}~2\pi)$} \\ \hline
$  1.961385  $&&$          1.961385$&\\
$ -2.532983$&$ +i\pi$&   $-2.532983$&$+i\pi$\\
$ -0.714134$&$ +i\pi/3$& $-0.705970 $&$+\pi/3$ \\ 
$ -0.714134$&$ -i\pi/3 $&$-0.705970 $&$-i\pi/3$ \\
$ -0.714134$&$ +i\pi$&   $-0.705970 $&$+i\pi$ \\
$  0.904667  $&&         $ 0.896503 $&$+i\pi/3$  \\
$  0.904667$&$ +i2\pi/3$ &$0.896503$&$ -i\pi/3$\\  
$  0.904667$&$ -i 2\pi/3$&$0.896503 $&$+i\pi$\\ \hline
$  1.961385   $&& $        1.961385$&\\
$ -2.532983$&$ +i\pi $&  $-2.532983$&$+i\pi$\\
$ -0.718967 $&&          $-0.742981$&\\
$ -0.718967 $&$+i2\pi/3$&$-0.742981$&$+i2\pi/3$\\ 
$ -0.718967 $&$-i2\pi/3 $&$-0.742981$&$-i2\pi/3$\\
$  0.909499 $&$+i\pi/3$&$   0.933514$&\\
$  0.909499 $&$-i\pi/3$&$   0.933514$&$+i2\pi/3$\\
$  0.909499 $&$ +i\pi $&$   0.933514$&$-i2\pi/3$\\ \hline
$  1.961385 $&&$            1.961385$&\\ 
$ -2.532983 $&$+i\pi$&$    -2.532983$&\\  
$  0.067492 $&$+i\pi/3$&$   0.095266 $&$  +0.757222i$\\ 
$  0.067492 $&$-i\pi/3 $&$  0.095266 $&$  -0.757222i$\\
$  0.067492 $&$+i\pi $&$    0.095266 $&$ + 1.337172i$\\
$  0.123040 $&$ +i2\pi/3$&$ 0.095266 $&$ - 1.337172i$\\
$  0.123040 $&$ -i2\pi/3$&$ 0.095266 $&$  +2.851617i$\\
$  0.123040$& &$            0.095266 $&$  -2.851617i$\\ \hline
\end{tabular}

\newpage

Table 10.

Examples of multiplets with $S^z_{\rm max}=5$ and $4$ with $L=16$ and
$\Delta=1/2.$ The roots listed are taken from the data for
$\Delta=-.501$ and we indicate by $\pm \infty$ roots which give
contributions to the total momentum of 
$\pm 2\pi/3$ in the limit $\Delta \rightarrow -1/2.$ We
give the other roots to 6 places but note that some of the values can
differ from their values at $\Delta=-1/2$ by as much as $0.07.$

\vspace{.3in}

Table 10a.

An example of the multiplet with $S^z=5,2,-1,-4$

\vspace{.1in}

$E=4.057\cdots,$ $P=2\pi/16$ 

\begin{tabular}{|rl|rl|rl|}\hline
\multicolumn{6}{|l|}{$S^z=5$}\\ \hline
$ -1.391351$&&&&&\\
$ -0.304001$&$+i\pi$&&&&\\
$ -\infty$&&&&&\\ \hline
\multicolumn{6}{|l|}{$S^z=2$}\\ \hline
 $-1.392407$&&$-1.391696$&&$ -1.391722  $& \\ 
 $-0.305603$&$ +i\pi$& $-0.302916$&$ +i\pi$&$ -0.302748$&$+i\pi$ \\  
$-\infty$&&$-\infty$&&$-\infty$&\\
$ -0.979344$&$+i\pi/3$&  $0.795702$&$+i2\pi/3$&$   0.793797$&$  +i\pi/3$ \\ 
$ -0.979344$&$ -i\pi/3$ &$0.795702$&$ -i2\pi/3$ &$ 0.793797$&$ -i\pi/3 $\\
$ -0.984361$&$ +i\pi$&   $0.795701$&&$           0.794323$&$ +i\pi$ \\ \hline
\multicolumn{6}{|l|}{$S^z=-1$} \\ \hline
$ -1.392093$&&$ -1.391968$&& $ -1.392068$& \\
$ -0.309847$&$ +i\pi$&$ -0.303059$&$ +i\pi$  &$ -0.301622$&$+i\pi$   \\
$ \infty$&&$ \infty$&&$ \infty$&\\
$\infty$&&$\infty$&&$\infty$&\\
$ -0.423815$&$ +i\pi/3$&$ -0.428616 $&$+i2\pi/3$ & $ 0.996443$&$ + i\pi/3$ \\
$ -0.423815$&$ -i\pi/3$&$ -0.428616$&$ -i2\pi/3$ &$  0.996443$&$  -i\pi/3$\\ 
$ -0.417366$&$+i\pi$&$    -0.428616 $&&$  1.000841  $&$+i\pi$\\ \hline
\multicolumn{6}{|l|}{$S^z=-4$}\\ \hline
$ -1.391504$&&&&&\\   
$ -0.302815 $&$+i\pi$&&&&\\
$ \infty$&&&&&\\
$\infty$&&&&&\\ \hline
\end{tabular}

\newpage

Table 10b.

An example of the multiplet with $S^z=5,2,-1$

\vspace{.1in}

$E= -4.167\cdots,$ $P=2\pi/16$ 

\begin{tabular}{|rl|rl|}\hline
\multicolumn{4}{|l|}{$S^z=5$}\\ \hline
$  0.251789$&&&\\
$  0.005266$&&&\\
$ -0.519473$&&&\\ \hline
\multicolumn{4}{|l|}{$S^z=2$}\\ \hline
$  0.251853$&& $ 0.251846$&   \\
$  0.005442$&&$  0.005500$& \\
$ -0.519607$&&$ -0.519683 $&\\
$  0.069044$&$+i2\pi/3$&$ 0.069117$&$ +i\pi/3$\\ 
$  0.069044$&$ -i2\pi/3$& $0.069117$&$  -i\pi/3$\\
$  0.069101 $&&$ 0.06898$&$  +i\pi$\\ \hline  
\multicolumn{4}{|l|}{$S^z=-1$}\\ \hline
$  0.251816$&&&\\
$  0.005524$&&&\\
$ -0.519555 $&&&\\
$ \infty$&&&\\
$\infty$&&&\\
$-\infty$&&&\\
$-\infty$&&&\\ \hline
\end{tabular}

\newpage
Table 10c.

An example of the multiplet with $S^z=4,1,-2$

\vspace{.1in}

$E= -5.478\cdots,$ $P=2\pi/16$ 

\begin{tabular}{|rl|rl|}\hline
\multicolumn{4}{|l|}{$S^z=4$}\\ \hline
  $0.721217$&&&\\
  $0.128539$&&&\\
 $-0.380667$&&&\\
 $-0.706916$&&&\\ \hline
\multicolumn{4}{|l|}{$S^z=1$}\\ \hline
$  0.721400 $&&$0. 721478$&\\    
$  0.128630 $&& $0. 128548$&\\ 
$ -0.380765 $&&$-0.380829$&\\    
$ -0.707118 $&&$-0.707220$&\\  
$  0.072111 $&& $ 0.072073$&$+i\pi$\\  
$  0.072175 $&$+i2\pi/3$&$ 0.072279$&$+i\pi/3$\\   
$  0.072175 $&$-i2\pi/3 $&$ 0.072279$&$ -i\pi/3$\\ \hline
\multicolumn{4}{|l|}{$S^z=-2$}\\ \hline
 $ 0.721274$&&&\\
 $ 0.128544$&&&\\
$ -0.380688$&&&\\
$ -0.706974$&&&\\
$ \infty $&&&\\
$-\infty $&&&\\ \hline
\end{tabular}}

\newpage
\vfill                 
\eject
\centerline{\epsfxsize=6in\epsfbox{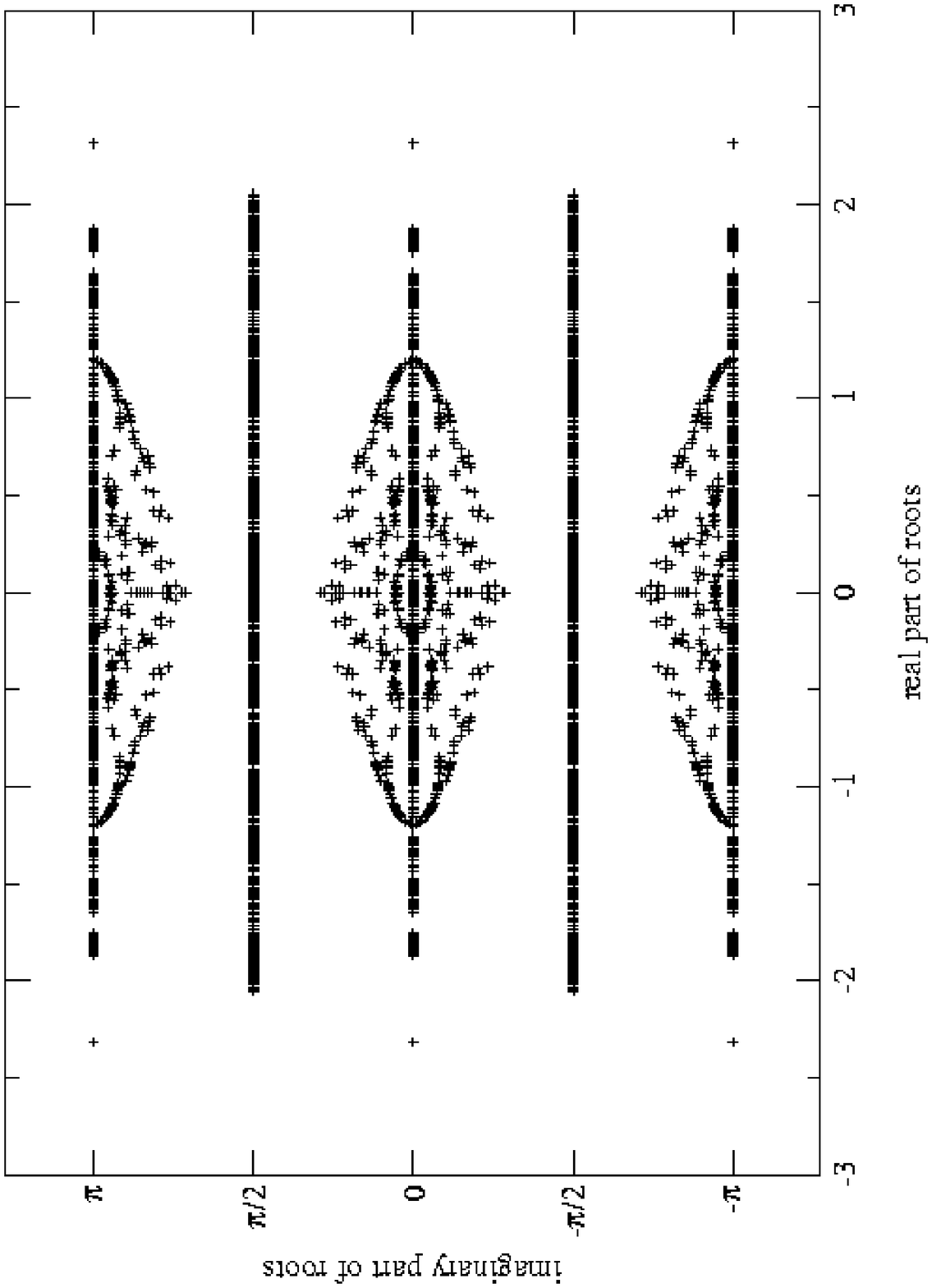}}

Fig. 1. Plot of the Bethe's roots for $\Delta=0,~S^z=0$ as obtained from
Baxter's exact expression\\
\centerline{ for $Q(v).$  The roots are symmetric both about the real axis 
and under the reflection $v\rightarrow -v.$}

\newpage
\vfill                 
\eject
\centerline{\epsfxsize=6in\epsfbox{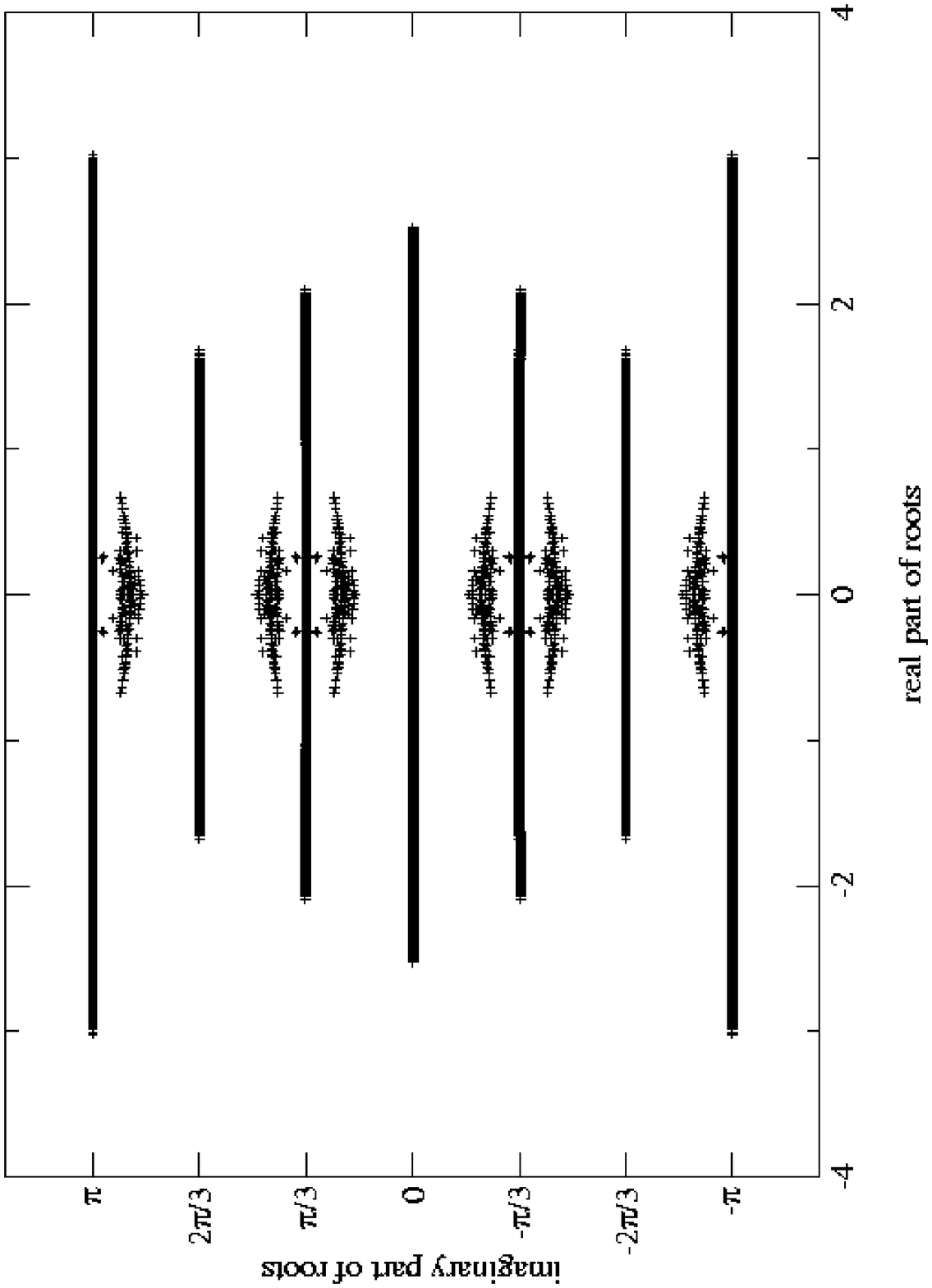}}

Fig. 2. Plot of the Bethe's roots for $\Delta=-1/2,~S^z=0$ as obtained from
Baxter's exact expression\\
\centerline{ for $Q(v).$ The roots are symmetric both about the real axis 
and under the reflection $v\rightarrow -v.$}

\end{document}